\documentclass[12pt]{article}
\usepackage{geometry,amsmath,graphicx,hyperref}
\newcommand{\Omicron}{\mathrm{O}}
\newcommand{\mathe}{\mathrm{e}}
\newcommand{\tmop}[1]{\ensuremath{\operatorname{#1}}}
\newcommand{\tmtextbf}[1]{{\bfseries{#1}}}
\newcommand{\tmtextit}[1]{{\itshape{#1}}}
\newcommand{\tmtexttt}[1]{{\ttfamily{#1}}}

\begin{document}

\title{Triviality of $\varphi^4$ theory in a finite volume scheme adapted to the
broken phase}
\author{Johannes Siefert and Ulli Wolff\thanks{
e-mail: uwolff@physik.hu-berlin.de} \\
Institut f\"ur Physik, Humboldt Universit\"at\\ 
Newtonstr. 15 \\ 
12489 Berlin, Germany
}
\date{}
\maketitle

\begin{abstract}
We study the standard one-component $\varphi^4$-theory in four dimensions.
A renormalized coupling is defined in a finite size renormalization scheme
which becomes the standard scheme of the broken phase for large volumes. 
Numerical simulations are reported using the worm algorithm in the limit
of infinite bare coupling. The cutoff dependence of the renormalized coupling
closely follows the perturbative Callan Symanzik equation and
the triviality scenario is hence further supported.
\end{abstract}

\begin{flushright} HU-EP-14/08 \end{flushright}
\begin{flushright} SFB/CCP-14-18 \end{flushright}
\thispagestyle{empty}
\newpage
\section{Introduction}

The Z(2) symmetric quantum field theory of a single scalar field with
$\varphi^4$ interaction is the number one textbook prototype model for all
kinds of field theoretic methods. At the same time, with its possibility of
spontaneous symmetry breaking, it may be viewed as the crudest caricature of
the Higgs sector of the Standard Model. In this context the strongly
conjectured triviality of the model in four space time dimensions is of
physical interest as it leads to upper bounds on the Higgs mass.\footnote{
We have to remark here that such bounds are not universal but
depend on the cutoff in use. Different lattice discretizations yield
different bounds, see \cite{Heller:1992pj} for example.
}
This is so because triviality means that the cutoff cannot be removed from the
regularized theory without ending in a free Gaussian theory. The model is
then interpreted as an effective theory in which an upper limit on tolerable
unphysical cutoff effects implies an upper bound on the interaction strength
which in turn is responsible for mass generation by the Higgs field.

Unfortunately
in the four dimensional case, we still have to rely on numerical methods to
demonstrate triviality beyond the perturbative regime. Such studies in the
lattice regularization have been strongly boosted by a series of papers by
L\"uscher and Weisz (LW), of which the first two have dealt with the one
component model in the symmetric {\cite{Luscher:1987ay}} and in the broken
{\cite{Luscher:1987ek}} phase. Here control over the lattice theory was gained
by combining large orders in the hopping parameter expansion with careful
perturbative renormalization group evolution.\cite{Brezin:1976bp} 
These studies were in addition
corroborated by some early Monte Carlo simulations as for example
{\cite{Montvay:1988uh}}, {\cite{Jansen:1988cw}}. In these cases the Ising
model was considered as the limit of $\varphi^4$ theory at infinite bare
coupling. Barring a complicated non-monotonic relation between bare and
(natural) renormalized couplings on the lattice, this limit is the most
interesting case for questions concerning triviality.

In recent years one of the authors has taken up the subject again after some
progress had been made in Monte Carlo methods which allow to achieve a new
level of precision in this context with only moderate investments in compute
power. The main new ingredients are on the one hand the use of so-called worm
algorithms {\cite{prokofev2001wacci}}, {\cite{Wolff:2008km}} to simulate
arbitrary order contributions of a hopping parameter expansion for observables
on finite lattices instead of generating field configurations. The second
ingredient is the use of finite volume renormalization schemes as in
{\cite{Luscher:1991wu}}. As triviality is an ultraviolet renormalization
effect, more compute power can be devoted in this way to closely approaching
the continuum limit as the thermodynamic limit does not have to be taken. In
other words, the manageable ratios $L / a$ between lattice size and spacing is
used to achieve a significant range of small $a$ and not for large $L$ in
physical units. In {\cite{Wolff:2009ke}}, {\cite{Weisz:2010xx}},
{\cite{Hogervorst:2011}} such a strategy has been explored for the symmetric
phase of the model. In this publication we now offer a finite size scaling
study on the other side of the critical line.

In section 2. we define our renormalization scheme, followed by basic
definitions of $\varphi^4$ theory. In section 4. the numerical method and
achieved results are described followed by a brief summary. This work is based
on the master thesis of the first author at Humboldt University, Berlin 2013.

\section{Broken phase finite volume scheme}\label{scheme}

At first glance the title of this subsection might look paradoxical as there
is no symmetry breaking in a finite volume. If we define however an order
parameter $v_0$ by the large distance behavior of the Z(2) symmetric
fundamental two point correlation ($\xi$ is the correlation length),
\begin{equation}
  \langle \varphi \left( x \right) \varphi \left( 0 \right) \rangle \cong
  v_0^2 \hspace{1em} \tmop{for} \hspace{1em} \left| x \left| \gg \xi, \right.
  \right.
\end{equation}
then this definition has a smooth thermodynamic limit. To define definite
renormalization conditions we employ the Fourier transform
\begin{equation}
  G \left( p \right) = a^4 \sum_x \mathe^{- i p x} \langle \varphi \left( x
  \right) \varphi \left( 0 \right) \rangle
\end{equation}
and extract $v_0^2$ from
\begin{equation}
  G \left( p \right) = L^4 \delta_{p, 0} v^2_0 + G_c \left( p \right),
  \hspace{1em} G_c \left( 0 \right) = 0,
\end{equation}
where we have assumed a torus of extent $L$ in each direction and $G_c$ is the
varying part of the correlation (`connected', although we here avoid the
one-point function).

We now complete our renormalization scheme by singling out two small torus
momenta
\begin{equation}
  p_{\ast} = \frac{2 \pi}{L} \left( 1, 0, 0, 0 \right), \hspace{1em} p_{\ast
  \ast} = \frac{2 \pi}{L} \left( 1, 1, 0, 0 \right)
\end{equation}
beside zero momentum. We match $G \left( p \right)$ to the form
\begin{equation}
  G \left( p \right) = Z \left\{ L^4 \delta_{p, 0} v^2 + \frac{1}{\hat{p}^2 +
  m^2} \right\}^{} \hspace{2em} \tmop{at} p \in \left\{ 0, p_{\ast}, p_{\ast
  \ast} \right\}
\end{equation}
which simultaneously fixes the wave function renormalization factor $Z$, a
renormalized expectation value $v$ and the renormalized mass $m$. By solving
these conditions we obtain
\begin{equation}
  z^2 = \left( m L \right)^2 = \frac{G \left( p_{\ast \ast} \right)
  \hat{p}^2_{\ast \ast} L^2 - G \left( p_{\ast} \right) \hat{p}^2_{\ast}
  L^2}{G \left( p_{\ast} \right) - G \left( p_{\ast \ast} \right)}
\end{equation}
and
\begin{equation}
  w^2 = \left( v L \right)^2 = \frac{G \left( 0 \right)}{G \left( p_{\ast}
  \right)}  \frac{1}{L^2 \hat{p}^2_{\ast} + z^2} - z^{- 2} .
\end{equation}
where we have introduced the dimensionless finite size scaling quantities $z$
and $w$ and the usual lattice momentum
\begin{equation}
  \hat{p}_{\mu} = \frac{2}{a} \sin \left( a p_{\mu} / 2 \right) .
\end{equation}

It is not difficult to see that in the thermodynamic limit $z \rightarrow
\infty$ our definitions of $m$ and $v$ approach those of $m_R$ and $v_R$ in
{\cite{Luscher:1987ek}}. Apart from this limit however, each fixed value of
$z$ defines a different renormalization scheme and the perturbative
coefficients of the continuum perturbative Callan Symanzik $\beta$ function,
for instance, will depend on $z$ beyond the scheme independent one and two
loop terms.

As usual in the spontaneously broken theory we define the renormalized
coupling constant in terms of $v$ by setting
\begin{equation}
  g = \frac{3 m^2}{v^2} = \frac{3 z^2}{w^2} . \label{gdef}
\end{equation}

\section{Some basic \texorpdfstring{$\varphi^4$}{phi**4} formulae}

The action in the lattice form is given by
\begin{equation}
  S = \sum_x \left[ \varphi \left( x \right)^2 + \lambda \left( \varphi \left(
  x \right)^2 - 1 \right)^2 \right] - 2 \kappa \sum_{x \mu} \varphi \left( x
  \right) \varphi \left( x + \hat{\mu} \right)
\end{equation}
with all dimensionless quantities. This is equivalent to the field theoretic
form
\begin{equation}
  S = a^4 \sum_x \left\{ \frac{1}{2} \left( \partial_{\mu} \phi \right)^2 +
  \frac{\mu_0^2}{2} \phi^2 + \frac{g_0}{4!} \phi^4 \right\}
\end{equation}
with mass dimension one field{\footnote{Our $\phi$ corresponds to $\varphi_0$
in the LW papers.}} $\phi$ if we match
\begin{eqnarray}
  a \phi & = & \sqrt{2 \kappa} \varphi \\
  a^2 \mu_0^2 & = & \frac{1 - 2 \lambda}{\kappa} - 8 \\
  g_0 & = & \frac{6 \lambda}{\kappa^2} = 6 \lambda \left( \frac{a^2 \mu_0^2 +
  8}{1 - 2 \lambda} \right)^2 . 
\end{eqnarray}
Classically, the symmetric phase arises for $a^2 \mu_0^2 > 0$ where $\varphi,
\phi$ fluctuate around zero with a bare mass given by
\begin{equation}
  m_0 = \mu_0 \hspace{1em} \left( \tmop{symmetric} \tmop{phase} \right) .
\end{equation}
For $a^2 \mu_0^2 < 0$ the field fluctuates around one of two equivalent
nonzero values $\pm \overline{\varphi}$. The quadratic fluctuations around
either constant field are now controlled by the bare mass
\begin{equation}
  m_0 = \sqrt{- 2 \mu_0^2} \hspace{1em} \left( \tmop{broken} \tmop{phase}
  \right) .
\end{equation}
Note that in LW $\mu_0$ does not appear, as in {\cite{Luscher:1987ay}} it is
replaced by $m_0$ while in {\cite{Luscher:1987ek}} the action in terms of
$\phi$ is not written and only the $m_0$ for the broken phase appears.
Consequently the relations between $m_0$ and $\kappa, \lambda$ differ in the
two papers as emphasized in a footnote in {\cite{Luscher:1987ek}}.

Following LW we explore the plane of bare parameters by approaching the
critical line (continuum limit) on trajectories at fixed $\lambda$ and define
the $\beta$-function
\begin{equation}
  \beta \left( a m, g \right) = \frac{\partial g}{\partial \ln \left( a m
  \right)} \left|_{\lambda} . \right.
\end{equation}
This definition entails the following tree level lattice artefact
contributions
\begin{equation}
  \beta \left( a m, g \right) = \frac{4 a^2 m^2}{8 + a^2 m^2} g + \Omicron
  \left( g^2 \right) \hspace{1em} \left( \tmop{symmetric} \tmop{phase} \right)
\end{equation}
and ($m^2 \rightarrow - m^2 / 2$)
\begin{equation}
  \beta \left( a m, g \right) = - \frac{4 a^2 m^2}{16 - a^2 m^2} g + \Omicron
  \left( g^2 \right) \hspace{1em} \left( \tmop{broken} \tmop{phase} \right) .
\end{equation}
Although these artefacts are small they may be avoided by switching to
modified couplings
\begin{equation}
  \tilde{g} = g \times \left\{ \begin{array}{ll}
    \left( 1 + a^2 m^2 / 8 \right)^{- 2} & \tmop{symmetric}\\
    \left( 1 - a^2 m^2 / 16 \right)^{- 2} & \tmop{broken}
  \end{array} \right. . \label{gtildedef}
\end{equation}

The perturbative continuum $\beta$ function for the coupling $g$ of the
previous subsection -- and also $\tilde{g}$ formed from it -- is given by
\begin{equation}
  \beta \left( 0, g \right) = b_1 g^2 + b_2 g^3 + b_{3, z} g^4 + \Omicron
  \left( g^5 \right), \hspace{1em} b_1 = \frac{3}{\left( 4 \pi \right)^2},
  \hspace{1em} b_2 = - \frac{17}{3 \left( 4 \pi \right)^4} \label{betaf}
\end{equation}
While the first two coefficients are scheme independent, the three loop term
$b_{2, z}$ is at the moment not known for our present scheme at finite $z$.
The infinite volume case is found in {\cite{Luscher:1987ek}},
\begin{equation}
  b_{3, \infty} = \frac{14.715616}{\left( 4 \pi \right)^6} . \label{beta2}
\end{equation}

\section{Worm simulations}

\subsection{Brief summary of the method}

The renormalization scheme of section \ref{scheme} is defined entirely in
terms of the two point correlation. Worm simulations are ideally suited for
its numerical computation in the Ising limit $\lambda = \infty$. The worm
ensemble is given by the partition function
\begin{equation}
  \mathcal{Z}= \sum_{u, v} Z \left( u, v \right) = \sum_{u, v, k} t^{\sum_{x,
  \mu} k \left( x, \mu \right)} \delta \left[ \partial_{\mu}^{\ast} k_{\mu} -
  q_{u, v} \right] . \label{Zens}
\end{equation}
In this formula we sum over link variables $k \left( x, \mu \right) \equiv
k_{\mu} \left( x \right) = 0, 1$ and $\delta \left[ \ldots \right]$ enforces
the constraint that is most easily described in words: each site except $u, v$
must be surrounded by an even number of $k = 1$ links while at $u, v$ 
(unless $u=v$) this
number must be odd. The fugacity is $t = \tanh \left(
2 \kappa \right)$. The $k$ configurations are in one-to-one correspondence
with strong coupling graphs with lines drawn on links with $k \left( x, \mu
\right) = 1$. At the same time we have the connection with the spin
formulation
\begin{equation}
  Z \left( u, v \right) =\mathcal{N} \sum_{\varphi} \mathe^{2 \kappa \sum_{x,
  \mu} \varphi \left( x \right) \varphi \left( x + \hat{\mu} \right)} \varphi
  \left( u \right) \varphi \left( v \right)
\end{equation}
where for the Ising limit the sum is over $\varphi \left( x \right) = \pm 1$
and $\mathcal{N}$ is a normalization factor. In {\cite{Wolff:2008km}} a lot
more details about this reformulation and the efficient simulation of
(\ref{Zens}) can be found. It is obvious now that the two point function is
given by
\begin{equation}
  \langle \varphi \left( x \right) \varphi \left( 0 \right) \rangle =
  \frac{\langle \langle \delta_{x, u - v} \rangle \rangle}{\langle \langle
  \delta_{u, v} \rangle \rangle}
\end{equation}
with the double angles referring to expectation values with respect to
(\ref{Zens}). The required Fourier transforms can be directly accumulated from
\begin{equation}
  G \left( p \right) = \frac{\langle \langle \mathe^{- i p \left( u - v
  \right)} \rangle \rangle}{\langle \langle \delta_{u, v} \rangle \rangle} =
  \frac{\left\langle \left\langle \prod_{\mu} \cos \left( p_{\mu} \left( u - v
  \right)_{\mu} \right) \right\rangle \right\rangle}{\langle \langle
  \delta_{u, v} \rangle \rangle} .
\end{equation}
For the last step we have used the invariance under individual reflections
along each direction. Note that with the small momenta of interest we do not
expect very rapid oscillations. As only ratios of $G$, where the wave function
renormalization cancels, are of interest, the denominator $\langle \langle
\delta_{u, v} \rangle \rangle = G \left( 0 \right)^{- 1}$ (inverse
susceptibility) is never really needed here.

\subsection{Numerical results}

\begin{figure}[htb!]
\centering
  \resizebox{0.8\textwidth}{!}{\includegraphics{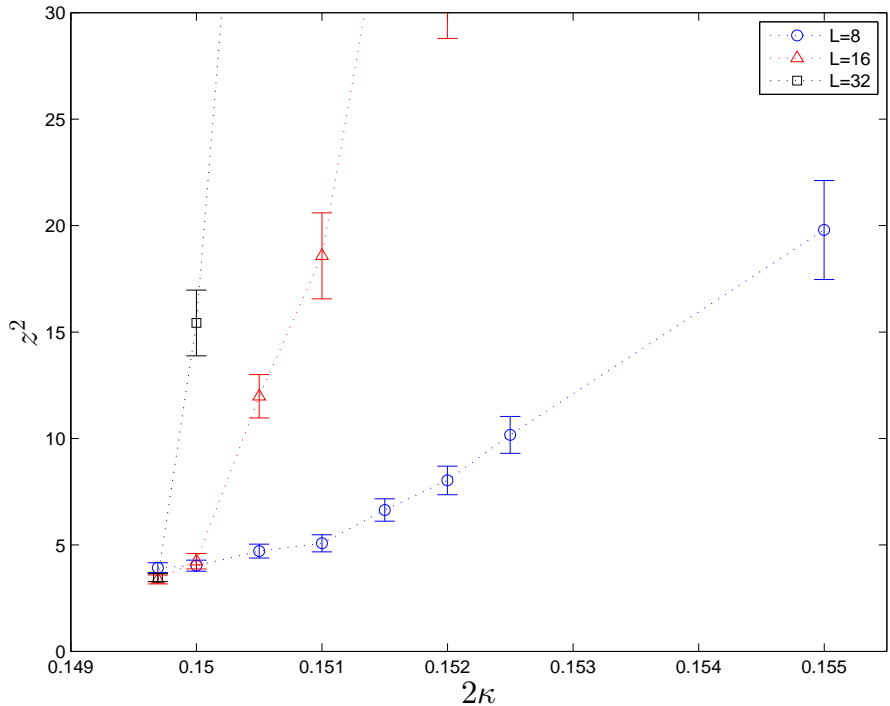}}
  \caption{Finite size mass against hopping parameter for $L/a = 8, 16,
  32$.\label{zkappa}}
\end{figure}

At first we have explored how $z$ depends on the hopping parameter for lattice
sizes $L / a = 8, 16, 32$. The results are shown in Fig.\ref{zkappa}. We are
here just above the infinite volume critical point which is known
{\cite{0305-4470-12-6-018}} to occur close to $2 \kappa \approx 0.149$. Each
data point in the plot corresponds to $10^6$ iterations, where an iteration
{\cite{Wolff:2008km}} consists of one worm move per site.
From these results we have decided to adopt in the following the target value
$z^2 = 10$ for our study. Our results are summarized in Table \ref{tabg}.
\begin{table}[htb!]
\centering
  \begin{tabular}{|l|l|l|l|l|}
    \hline
    $L$ & $2 \kappa$ & $z^2$ & $g$ & $\tilde{g} \left|_{z^2 = 10} \right.$\\
    \hline
    {\phantom{1}}8 & 0.152460 & 10.024(96) & 29.13(30) & 29.70(26)\\
    \hline
    12 & 0.150992 & 10.008(98) & 24.88(26) & 25.09(22)\\
    \hline
    16 & 0.150450 & {\phantom{1}}9.964(99) & 22.39(24) & 22.51(20)\\
    \hline
    24 & 0.150046 & {\phantom{1}}9.974(98) & 19.65(21) & 19.70(18)\\
    \hline
    32 & 0.149899 & {\phantom{1}}9.980(97) & 17.95(19) & 17.97(16)\\
    \hline
    48 & 0.149790 & 10.065(96) & 15.90(17) & 15.89(14)\\
    \hline
  \end{tabular}
  \caption{Simulation results to determine the renormalized coupling in the
  continuum limit (growing $L / a$) for fixed $z^2 = 10$.\label{tabg}}
\end{table}
Each line corresponds to a statistics of $8 \times 10^7$ iterations. By some
tuning we found values of $\kappa$ that lead to $z^2 = 10$ within errors.
The directly measured couplings (\ref{gdef}) are given in the fourth column
while the rightmost column differs by two tiny corrections. By the first order
reweighting technique described in {\cite{Wolff:2009ke}} the value is adjusted
to $z^2 = 10$ exactly and then the cutoff correction (\ref{gtildedef}) is
applied. The first correction is clearly only a change within the error bars,
but, although to a much lesser degree than in {\cite{Wolff:2009ke}}, it in
addition lowers the statistical error slightly.
\begin{figure}[htb!]
\centering
  \resizebox{0.8\textwidth}{!}{\includegraphics{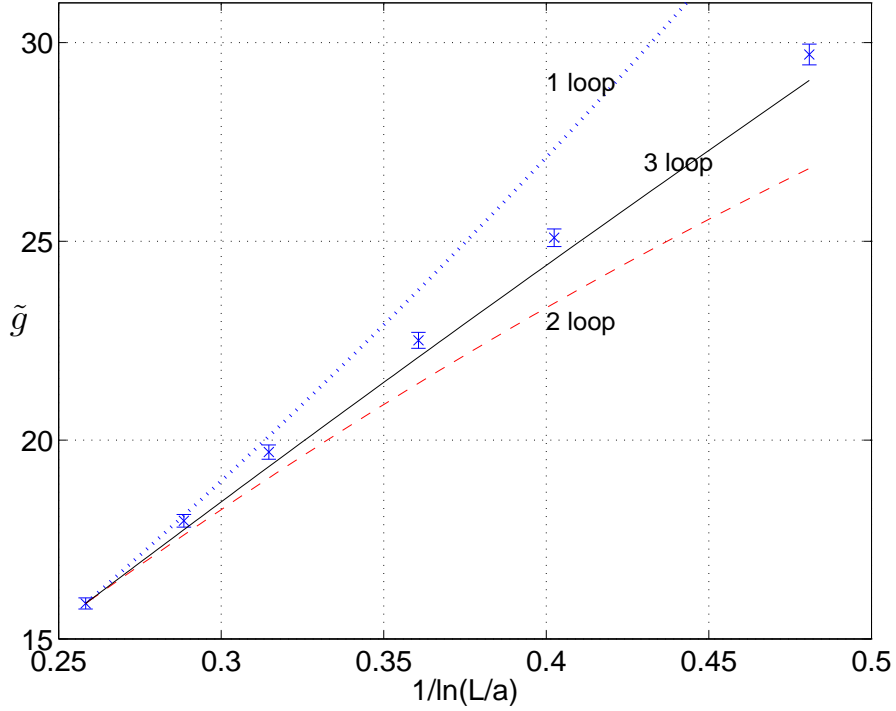}}
  \caption{Coupling $\tilde{g}$ at $z^2 = 10$ as a function of the cutoff. The
  curves stem from integrations of the renormalization group equation at
  various loop order truncations. The leftmost point is taken as initial
  value. \label{figg}}
\end{figure}
These data are plotted in Fig.\ref{figg}. The dotted (blue), dashed (red) and
solid (black) curves derive from integrating the Callan Symanzik equation
\begin{equation}
  \frac{d \tilde{g}}{d \ln \left( L / a \right)} = - \beta \left( 0, g \right)
\end{equation}
with the continuum $\beta$-function at 1,2,3 loop perturbative precision.
Beside the universal coefficients (\ref{betaf}) we here use the infinite
volume value (\ref{beta2}) for the three loop coefficient. As discussed before
this is only indicative with the presently unknown coefficient for $z^2 = 10$
certainly being slightly different. The experience in the symmetric phase has
been, however, that at this size the difference may not be very sizeable.

\section{Summary}

We have defined a finite size renormalization scheme for $\varphi^4$ theory,
which in the infinite volume limit goes over into the one that is standard in
the broken phase of the model. In the Ising limit of infinite bare coupling,
we have numerically generated values of the renormalized coupling as a
function of the lattice cutoff. Using novel simulation techniques we computed
precise values which turn out to closely follow the perturbative
renormalization group. The data points are nicely sandwiched between the one
and two loop trajectories. The three loop curve falls in between and is only
about two sigma (2 \%) away from our data, although the three loop coefficient
is taken for $z^2 = \infty$ rather than $z^2 = 10$ studied here. If we
conclude agreement with perturbation theory in the range studied then this
should be even better justified for larger $L / a$ and $\tilde{g}$ tends to
zero in the continuum limit at a logarithmic rate. This supports the
triviality scenario once more by combining numerical and perturbative methods.

\section*{Acknowledgements}

We would like to thank Tomasz Korzec and Peter Weisz for very helpful
discussions and comments on the manuscript.

\end{document}